\documentstyle[eqsecnum,aps,twocolumn]{revtex}
\begin{document}
\small
\textwidth 7.0in
\def\baselinestretch {0.7}
\setcounter{section}{0}
\setlength{\parskip}{0 pt}
\title{Correct interpretation of trace normalized density matrices as ensembles}
\author{Paul M. Sheldon}
\date{22 June 1996}
\maketitle
\begin{abstract}
A density operator,  $\rho \rm \ =\ {P}_{\alpha }\ |\alpha \rm >\ <\alpha \rm |\ +\ {P}_{\beta }\ |\beta \rm >\ <\beta \rm |$,  with ${P}_{\alpha }$ and ${P}_{\beta }$ linearly independent normalized wave functions,  must be traced normalized, so ${P}_{\beta }\ =\ 1\ -\ {P}_{\alpha }$. However, unless $<\alpha \rm |\beta \rm >\ =\ 0$, ${P}_{\alpha }$ and ${P}_{\beta }$ cannot be interpreted as probabilities of finding $|\alpha \rm >$ and $|\beta \rm >$ respectively. 
We show that a density matrix comprised of two (${P}_{\alpha }$ and ${P}_{\beta }$  nonzero) non-orthogonal projectors have unique spectral decomposition into diagonal form with orthogonal projectors. Only  then, according to axioms of Von Neumann and Fock,  can we have probability interpretation of that density matrix,  only then can the diagonal elements be interpreted as probabilities of an ensemble.
 Those probabilities on the diagonal are not ${P}_{\alpha }$ and ${P}_{\beta }$.
 Further,  only in the case of orthogonal projectors can we have the degenerate situation in which multiple ensembles are permitted.
\end{abstract}
\section{Introduction :}

This paper is an attempt to elucidate the well known Schrodinger cat paradox with reference to its debate in Sir Roger Penrose and Stephen HawkingÕs recent joint book, ``The Nature of Space and Time'', by more fully presenting decoherenceÕs view that the wave function has merely an \underline {interpretation} in probability rather than a  \underline {definition} in probability.

Even though, throughout the text, Penrose and Hawking publicly debate their views on the paradox, the casual reader could mistakenly conclude that just one page of PenroseÕs algebra, page 70, effectively evaluates and dismisses the promising young program of decoherence.

Penrose uses both wave function and density matrix algebraic representation of the so called correlated state of SchrodingerÕs cat coupled to a memory state. Penrose claims decoherence doesnÕt explain the predicament of SchrodingerÕs cat because decoherence does not select from what he \underline {defines} as two alternative ensembles. 

Penrose postulates that these alternative ensembles are identical to weighted sums of wave function or ketbra products. But, as Zeh says (page 2 of his October 96 quant-ph entry), these are not ensembles at all : 
\begin{quotation}
This density matrix has a form {\it as though} it represented an ensemble of wave functions, ${\phi
}_{n}^{local}$, with respective probabilities,
${p}_{n}$. Equation (2) demonstrates, however,
that it does not. This confusion between proper
and ``improper" mixtures has given rise to the
most frequent misinterpretation of decoherence as
leading to ensembles, and thus as deriving the
collapse of the wave function as a stochastic
process.
\end{quotation}

Decoherence never claimed to select between the things that Penrose calls ensembles.

PenroseÕs ``ensemble'' 1 is [cat dead] x [observed dead memory state] or [cat alive] x [observed alive memory state] presumably orthogonal and without any overlap and ``ensemble'' 2 is two different superpositions of dead and alive (in products of observed and recorded state) that add up to the same thing in both state and density matrix representations. 

Postulating this, doesnÕt lead to superselection.

Penrose, in the book, represents decoherence as an algebraic recognition about ways to arrange sums of products of factors. But, only a dynamical process will truly represent it. It is not, as in algebra, instanteous, but instead happening over time (decoherence time). Decoherence only derives its power from the physical interaction between those factors, an interaction not addressed in PenroseÕs algebra.

We next consider the physical meaning of the Schmidt polar form of orthogonal states, a purely formal device which predates any notions of quantum mechanics and certainly doesnÕt deal with the essential environmental interaction in the extended Von Neumann measurement model.

Penrose seems to imply that the orthogonality imposed by that form both is decoherenceÕs only claim to explanatory power and weakens the generality.

Regardless of measurement interaction, at each instant, one can find a sum of orthogonal wave function products called a Schmidt polar form. This, in itself, represents neither restriction (information about physical law or principle) nor loss of generality.

While the Schmidt polar form together with the orthogonality of the products will be required for superselection according to decoherence, that form alone isnÕt sufficient for the apparent gain of information by an experiment.

Rather, as Albrecht showed, for a sufficiently ``large'' measurement apparatus, \underline {interaction} would, after a time, produce dynamically stable (to repeated measurement interaction) orthogonal states in a Schmidt polar form to counter the instability caused by \underline {nearly} degenerate density matrix eigenvalues (probabilities). 

Penrose sets up an extreme situation of \underline {precisely} denerate eigenvalues on p.70, the interpreted probabilities being mathematically prepared as exactly equal. 

I claim it is wrong to imagine oneself able to so determine an initial wave function. 

To see measurement as a collapse into a definite outcome with a certain probability requires that a repetition get that same outcome.

I first consider two systems in quantum mechanics. I let one system be an observerÕs memory of an experiment and the other system be what is being observed. Quantum mechanics not only allows superposition, but the dynamics of the wave equation demands it.  So we not only \underline {may} consider the so called correlated state of these two systems (the superposition of the product pairs of a state from the first and a state from the second) but we \underline {must} consider it. 

In fact, according to Wheeler and Everett who addressed the crisis of interpretation produced by quantum cosmology in the absence of external observer, etc.), we must deal with several copies of observers observing different universes. 

The unextended Von Neumann model (discussed in the last two pages of his 1930 monograph) already explained the unitary dynamics of measurement.

To impose time asymmetry, Von Neumann starts with a special initial state, a single product. He then shows, the measurement interaction will make that single product ``branch'' into a superposition of products. But, which products, Penrose might ask?

The function subspace of the observed system can be broken up into a special orthogonal decomposition that corresponds to the eigenvalues of a Hermitian operator, an observable.

A measurement apparatus, though its design was later recognized to be curtailed by the environment, must measure some such eigenvalue. Associated with the apparatus, according to Von Neumann, is an interaction Hamiltonian, comprised of two factors, one from the observed system, and one from the pointer. The former, with the eigenvalue we want to observe, is multiplied by the momentum of the pointer. The interaction Hamiltonian, suitably placed in an exponent, creates a time-evolution operator in the so-called interaction picture. For each eigenfunction building the observed system, this operator simplifies to an excursion operator on the pointer. And it distributes linearly over the system pointer wavefunction to produce what Zurek calls only the first stage of measurement, the correlated state. This correlated state is a superposition of products of an eigenfunction of the system factor and eigenfunction of the pointer position. 

This ``first-stage'' correlated state can be rewritten algebraically to reflect alternative superpositions. That is why, Zurek says, it is incomplete and only the first stage.

The next stage occurs when you ``record'' the
response of the pointer into your history and it
\underline {remains consistently recorded},
something that essentially involves the
huge-dimensioned \underline {environment} outside
the pointer and measured system. This is the
second stage of measurement, \underline
{decoherence}. 

You are looking at the state multiplied by that pointer value in the correlated state sum, because (from an ``outside'' multiverse point of view, a point of \underline {view} that paradoxically doesnÕt exist but we talk about anyway) ÒyouÓ are there in one of the products to view the other factor. Environmental decoherence doesnÕt allow a viewpoint where you see correlated states for ``macroscopic'' things which are very entangled with the environment. Instead it selects a Schmidt polar form where all observers that see each other get on the same product and agree on the recorded history in that product. Communicate with any of WignerÕs friends and you will agree on a measurement result.

Relative to one particular you  something happened, a collapse. The features of collapse are not only a moment by moment probability but also a permanence of record of what happened.

Relative to the multiverse nothing really happened but everything just is ``there'', the ``old'' you splitting relative to some clock pointer position onto many products. 

So, relative to you, \underline {before} you record the response of the pointer, you must have something like a probability of what you would record later. What allows the interpretation of this probability-like thing as a probability?

Trace out the environmental and system degrees of freedom from the full density matrix of the pure state of the universe and you will get a reduced density matrix of the pointer. This reduced density matrix is trace normalized and can be orthogonalized to have its positive definite eigenvalues interpreted as probabilities of pointer positions. These probabilities turn out exactly the same as those associated with system states. How can this happen?

Expectations in the above so called improper mixtures have identical formal appearance to those in proper mixtures and the identical formal appearance of expectations lead us to speak of probabilities in ``improper'' circumstances.

The reduced density matrix can be used to define expectation values of operators, exactly as one built from real probabilities can. Not only that, consider finding the expectation of a special class of operators, the orthogonal projection ketbras.

These special expectations can define the probability of distinct eigenvalues of a Hermitian operator. Consider the spectral decomposition of that operator into projections times eigenvalues. The expectation value operator is linear, so that it can not only be distributed, but also pulled all the way out onto the individual projections. Now one sees that this expectation value of the operator conforms to the ordinary probability interpretation of expectation value of eigenvalues times probabilities of eigenstate. 

In addition, for a probability interpretation of the outcome of a measurement, a repeated measurement must yield a certitude.

The eigenvalues of the reduced density matrix of the pointer are what are interpreted as probabilities of the observable's operator if the environmentÕs measurement-like interaction doesnÕt change the observableÕs eigenvalue, i.e. if its measurement-like interaction is a quantum non-demolition one relative to the observableÕs operator.

Whether or not the eigenvalues of the Hermitian density matrix of the pointer are degenerate, we can always construct an orthogonal set of eigenvectors.

What about other orthogonalities? By definition of Von Neumann measurement interaction with Hermetian operator, the dynamics of branching resolves orthogonal system states or subspaces. These distribute themselves among the terms of the correlated state produced by the excursion operator. Now consider other factors in each of these terms, namely the state of the observer, bath (environment), and discrete pointer position. These factors are only after decoherence time almost orthogonal. Albrecht provides and example showing that when one has a very large-dimensional apparatus, one can afford to have pointer probabilities closer to each other and still accept that the states produced by the excursion operator are close to providing the required orthogonal states of the Schmidt polar form.

Why then is this dimension enormous and how does this enormity allow almost orthogonality rather than the ambiguities of ZurekÕs so-called first stage of measurement?

The physical pointerÕs degrees of freedom are really combined with the environmentÕs degrees of freedom, e.g. photons that scatter off it. The photons are producing correlated states with the physical pointer and thus are themselves engaged in a measurement interaction along with the observer correlating to the photons.

In dealing with a random measurement interaction , Albrecht points out that one large dimensioned unit vector randomly kicked by measurement around the unit sphere will have probably negligible projection on any one arbitrary axis and hence is probably almost orthogonal to that arbitrary axis. So, augmented by such a random environmental interaction, and after a so called decoherence time, which allows the dimension to get large enough, the Òobserver statesÓ (pointer and environment) are almost orthogonal in a Schmidt polar decomposition of product pairs of orthogonal states.

The above example with complete randomness in the environment interaction makes plausible this almost orthogonality in a simple extreme way. When that interaction is not completely random, Zeh points out that one has examples of superselection. This means that certain things we propose to measure with an interaction might be erased by the environment.

Take, for example, macroscopic parity nonconservation. Due to the enormous interaction ÒsizeÓ with an environment of chirality-measuring photon polarization states, which erase parity, we had best not set our pointer to measure parity!  On the other hand, measuring chirality repeatedly yields idempotency (along a branch) and thus can be interpreted as having recorded a certitude where there had been an ensemble. One then has a consistent history! Decoherence thus explains the puzzling fact that sugars are found in definite chiralities.

There are many other examples of such superselections which are well explained by decoherence.

Using only decoherenceÕs ``traceout'' feature to reduce the density matrix will fail to superselect between those ensembles which are supposedly more ÒgenerallyÓ defined by Penrose et al. By using merely this feature as generating the probability, the casual reader of PenroseÕs position might be misled to believe all the physics can be encrypted in an arbitrary and more general reduced density matrix. 

I will illustrate, at first, with PenroseÕs simple two-state system and memory, that his ÒensemblesÓ arenÕt really more general. Penrose has implied this greater generality with the words ``\underline {not necessarily} orthogonal'' in his final retort to Hawking on p.134. A referee of one of my recent papers had refuted this when he stipulated orthogonality as an obviously required condition for ``interpretation as'' (as distinct from ``being'') probabilities.

Penrose cites a paper on ostensibly more general rho-ensembles, which ignores physical interaction dynamics, in contrast to proponents of decoherence, including the above mentioned referee. 

By using PenroseÕs postulate of what an ensemble is before decoherence-time rather than talking about something else that is interpreted as an ensemble after decoherence-time, we and not nature must choose the eigenensemble.

I will illustrate here that we can WOLOG use ``eigenensembles''.

In fact, I also propose, in thesis, to study what else might be necessary ``on the way to sufficiency'' for probability interpretation. I use this phrase because probability interpretation (measurement resolvability of a superselected ensemble) is something only approached approximately in a so-called (e folding) ``decoherence time''.

This physics dealing with time is not encrypted in the reduced density matrix! The reduced density matrix is not an ensemble. The Von Neumann operational definition of measurement and formally an infinite amount of time allows us to interpret what is not an ensemble as an ensemble and to see a special eigenensemble as superselected. 

That is what the decoherence research has been illustrating in recent years and what  seems not to be illustrated,  but rather misrepresented, by PenroseÕs algebra on p.70.

A finite time of apparatus-environment interaction only fails to superselect between eigenensembles that are too close to each other. ``Closeness'' of ensembles involves two factors, the smallness of the difference of eigenvalues of the reduced density matrix (eigenensemble probabilities) and also the size of the measuring apparatus-environment (entanglement dimensionality that increases in time with each branching due to another measurement).

``Buying a theory'' is motivated by both cost (complication) and payoff (use). Buying decoherence is no different.

When a person ``buys'' a theory or way of interpreting things, he pays what I call OccamÕs cost of complication. OccamÕs razor would have us choose the interpretation of greatest simplicity.

OccamÕs razor motivates our choice of a theory by its cost alone. The cost of decoherence is the introduction of unobservable universes (using only observed and tested quantum dynamics) while the cost of collapse is more laws (to cut out extra universes). OccamÕs razor alone canÕt choose between them.

The advocate of decoherence, however, gets a payoff. He may design his instrument to either resolve or blur close by eigenensembles. He can blur by looking at things before decoherence time or resolve by looking after. The collapse conjecture of Penrose doesnÕt seem to have yet predicted any such time with hard numbers like Zurek has.

If the advocate of decoherence still wants to see these alternate ensembles after a time of entanglement (e.g. quantum computers after the required number of machine cycles), he designs that apparatus to remain in that time of small size by isolating it from the environment interactions, for example, by refrigeration. Then, the decoherence time limit will allow the decoders enough quantum computer machine cycles to solve the cryptographic factoring problem.

If the advocate of decoherence wants to rapidly see whether SchrodingerÕs cat is alive or dead, he lets the apparatus get large in the limited time allowed by the experimental schedule.

To have similar design options, an advocate of collapse must yet find a collapse time where Zurek and others have already found a decoherence time. How easy will that be to do?

Consider some excerpts from PenroseÕs ÒThe EmperorÕs New MindÓ (on p.367, 368, 369 and 372) :
\begin{quotation}

\ldots

At some stage, this complex quantum linear superposition becomes a real probability-weighted collection of actual alternatives ... According to the viewpoint that I am proposing, that stage occurs as soon as the difference between the gravitational fields of the various alternatives reaches the one-graviton level.

\ldots

How `big' is the one graviton level? \ldots more a
question of mass and energy distribution. \ldots
The characteristic quantum-gravitational scale of
mass is what is known as the Planck mass \ldots
These must be allowed to be what are called
longitudinal gravitons - the ÒvirtualÓ gravitons
that compose a static gravitational field.
Unfortunately, there are theoretical problems
involved in defining such things \ldots .

\ldots

(Italics are mine)
\end{quotation}

I brief the above :

To determine a time of collapse, one must find a method of counting to a very obscure {\bf one}. For, collapse will evidently happen when there is enough curvature difference between would-be superposed states, a difference of environmental curvature related vaguely to {\bf one} (virtual and therefore possibly massive) exchange-force-graviton of about Planck-mass. 

In contrast to predicted and therefore experimentally testable decoherence time in linear quantum mechanics, where a size is found by merely counting the dimensions of an enlarging apparatus-environment, evidently a collapse time will be for quite some time immune to verification.

On an interesting side note, H. D. Zeh pointed out to me a publication by Kiefer, a decoherence advocate. Kiefer treated therein real gravitons causing decoherence. Treating vacuum solution minisuperspaces (oneÕs that only contain gravity coupled to nothing else), Kiefer attempted to show therein that real gravitons cause decoherence, superselect them as classical and render a classical, though so-called internal, time.

John Wheeler, an advocate of decoherence, pointed out in Batelle Rencontres the crisis in {\bf interpretation} of quantum mechanics brought out by the new field of quantum cosmology, the absence of an external observer (and external time). 

Yet, an advocate of collapse would sell us a complication of theory at what would be the final (somehow macroscopic) point in the correlation chain, consciousness (better than any currently understood computer). Here he would demand an extended quantum theory that gives actual ensemble states and not a single consciousness correlated state. He would demand {\bf not an interpreting}, participating, branching consciousness, but one single {\bf external} consciousness, perhaps an isolated quantum computer that could observe.

Since any one really correlated consciousness would see one reality or another, what difference is there to saying, by a sort of ``relative-to-me'' principle, that the other realities arenÕt there? 

The difference isnÕt seen by OccamÕs razor. 

But an advocate of decoherence has the payoff that he can recognize that an isolated quantum computer not isolated from the observation process is a contradiction in terms. I maintain that the collapse advocate puts together too much in one system. He puts together the time irreversible recording of computational results with what Fredkin and others have shown to be a  fundamentally time reversible quantum computational process.

So, rather than classical collapse and ``relative-to-me'' possibilities being the ultimate arbiter of what is, we might, instead, through quantum principles design things as we please, the classical or nonclassical, computer reversible process or irreversible record, and perhaps even time or nontime.

A recent Texas Instruments meeting affirmed that the power to specify options like this goes beyond ``mere'' physics curiosity.

Seth Lloyd, a mechanical engineer at MIT, demonstrated (Science 17 Sept 1993, Vol.261, pp.1569-1571) that coherent Pi-pulses can both enter data into, and manipulate, quantum two state systems (so called q-bit registers).

In discussion following the meeting, we agreed in the usefulness for these computers of my proposed study of superselected states and the derivation and design of their classicality. He maintained that his coherent Pi-pulses, because they are most classical, are most suited as ``probes'' of (as distinct from measurements in) quantum mechanics. For clarity, letÕs call such a probe a ``q-observer'' to distinguish it from an observer that records measurements. We see only the first stage of measurement!  How is this so?

I had already followed the mathematics of a seminal paper, (Zurek, Physics of Time Asymmetry,  ``Preferred sets of states, predictability, classicality, and environment-induced decoherence'' p.175-212) and found that coherent state classicality comes from having the least Von Neumann entropy production, the least decoherence. That is, we might say classicality almost happens when decoherence almost stops. Ironically, that also implies the least disturbance of what is left of the quantum mechanical. 

Seth Lloyd and I both realized that this minimum Von Neumann entropy production of a superselected classical state could have a dual role to play.

We cited CavesÕ work with density matrices and mutual information considerations and saw that, were a q-observer and the q-observed designed in isolated interaction starting from an initial pure state, their reduced density matrices would show identical entropies. Minimum Von Neumann entropy production by a classical state would guarantee the same for the state which it probes. 

So, the dual roles are first for the state itself and second for a quantum mechanical thing with which it alone interacts. 

While a most classical state, say a coherent Pi-pulse, is the least measured i.e. decohered by the environment, it also measures the least i.e. decoheres a quantum computer the least, thus disturbing the benefits of their quantum superposition i.e. parallelism the least. 

Such classical (collapsed as much as possible) states have a criterion for payoff, a criterion based on a predictable time limit (for quantum computing within), namely the decoherence time.  Decoherence fulfills the Popper requirement of a theory, providing predictions that can be falsified by experiment. If that decoherence time consistently proves to be a criterion of payoff, that would weigh heavily for decoherence theory rather than collapse theory which offers no such payoff.

So, I would study generalized coherent states to better understand what is really going on with classicality for more reliable quantum computer probes. 
\section{Analysis :}
\begin{samepage}
 Without loss of generality of projector decomposition,  we let phase of $|\alpha \rm >$ and $|\beta \rm >$ such that $<\alpha \rm |\beta \rm >$ is real and positive :
\begin{equation}
<\alpha \rm |\beta \rm >\ =\ \left|{<\alpha \rm |\beta \rm >}\right|\ =\ c\ \ne \ 0
\end{equation}
\end{samepage}

\begin{samepage}
 Now,  because $|\alpha \rm >$ and $|\beta \rm >$ are linearly independent,  each eigenvector may be expressed as a unique linear combination of $|\alpha \rm >$ and $|\beta \rm >$. Let us express  normalized eigenvectors, $|{\widehat{e}}_{1}>$ and $|{\widehat{e}}_{2}>$, in terms of $|\alpha \rm >$ and $|\beta \rm >$  :
\begin{equation}
\matrix{|{\widehat{e}}_{1}>\ =\ {c}_{1\alpha }\ |\alpha \rm >\ +\ {c}_{1\beta }\ |\beta \rm >\cr
|{\widehat{e}}_{2}>\ =\ {c}_{2\alpha }\ |\alpha \rm >\ +\ {c}_{2\beta }\ |\beta \rm >\cr}
\end{equation}
\end{samepage}

\begin{samepage}
We will now investigate a proposition, that one of $|\alpha \rm >$ and $|\beta \rm >$ appears in both linear combinations for the eigenvectors. We suppose the contrary and will achieve a contradiction. 

If ${c}_{1\beta }\ =\ {c}_{2\alpha }\ =\ 0$, then each of $|\alpha \rm >$ and $|\beta \rm >$ appears in only one linear combination. Then we may as well write :
\begin{equation}
\matrix{|{\widehat{e}}_{1}>\ =\ |\alpha \rm >\cr
|{\widehat{e}}_{2}>\ =\ |\beta \rm >\cr}
\end{equation}
\end{samepage}

\begin{samepage}
Applying the density operator to $|{\widehat{e}}_{1}>$, we find  :
\begin{equation}
\matrix{{P}_{{\widehat{e}}_{1}}|\alpha \rm >\cr
=\ {P}_{{\widehat{e}}_{1}}|\ {\widehat{e}}_{1}>\cr
=\ \rho \rm \ |\ {\widehat{e}}_{1}>\ \cr
=\ \left({{P}_{\alpha }\ |\alpha \rm >\ <\alpha \rm |\ +\ {P}_{\beta }\ |\beta \rm >\ <\beta \rm |}\right)|\alpha \rm >\cr
=\ {P}_{\alpha }\ |\alpha \rm >\ <\alpha \rm |\alpha \rm >\ +\ {P}_{\beta }\ |\beta \rm >\ <\beta \rm |\alpha \rm >\cr
=\ {P}_{\alpha }\ |\alpha \rm >\ +\ c\ {P}_{\beta }\ |\beta \rm >\cr
\Rightarrow \rm \ {P}_{\beta }\ =\ 0\cr}
\end{equation}
\end{samepage}

\begin{samepage}
But, this contradicts what we had assumed.

Therefore, one of $\left\{{|{\widehat{e}}_{1}>,\ |{\widehat{e}}_{2}>}\right\}$ appears in both linear combinations for eigenvectors (one of ${c}_{1\beta }\ or\ {c}_{2\alpha }\ \ne \ 0$) and WOLOG we relabel and make that be $|\alpha \rm >$. We can then simply, with a suitable choice of phase,  write a system of unnormalized eigenvectors where the coefficiennt of  $|\alpha \rm >$ is one and ${r}_{j}$ is defined to make the coefficient of the other be $c\ {r}_{j}$. Our system of unnormalized eigenvectors becomes :
\begin{equation}
\matrix{|{e}_{j}>\ =\ |\alpha \rm >\ +\ c\ {r}_{j}\ |\beta \rm >\cr
where\ |{\widehat{e}}_{j}>\ =\ {|{e}_{j}> \over \sqrt {<{e}_{j}|{e}_{j}>}}\cr}
\end{equation}
\end{samepage}

\begin{samepage}
Surpressing the index j, we tabulate some results for later use :
\begin{equation}
\matrix{<\alpha \rm |e>\ =\ <\alpha \rm |\alpha \rm >\ +\ c\ r\ <\alpha \rm |\beta \rm >\ =\ 1\ +\ {c}^{2}\ r\cr
<\beta \rm |e>\ =\ <\beta \rm |\alpha \rm >\ +\ c\ r\ <\beta \rm |\beta \rm >\ =\ c\ \left({1\ +\ r}\right)\cr}
\end{equation}
\end{samepage}

\begin{samepage}
 Imposing the  condition that $|e>$ is an eigenvector of $\rho $, we find that :
\begin{equation}
\matrix{{P}_{e}\ \left({|\alpha \rm >\ +\ c\ r\ |\beta \rm >}\right)\ \cr
=\ {P}_{e}\ |e>\ =\ \rho \rm \ |e>\ =\ \left({{P}_{\alpha }\ |\alpha \rm >\ <\alpha \rm |\ +\ {P}_{\beta }\ |\beta \rm >\ <\beta \rm |}\right)|e>\cr
=\ {P}_{\alpha }\ |\alpha \rm >\ <\alpha \rm |e>\ +\ {P}_{\beta }\ |\beta \rm >\ <\beta \rm |e>\cr
=\ {P}_{\alpha }\ |\alpha \rm >\ \left({1\ +\ {c}^{2}\ r}\right)\ +\ {P}_{\beta }\ |\beta \rm >\ c\ \left({1\ +\ r}\right)\cr}
\end{equation}
\end{samepage}

\begin{samepage}
By equating above coefficients of linearly independent $|\alpha \rm >$ and $|\beta \rm >$, we obtain a system :
\begin{equation}
\matrix{{P}_{e}\ =\ {P}_{\alpha }\ \left({1\ +\ {c}^{2}\ r}\right)\cr
{P}_{e}\ c\ r\ =\ {P}_{\beta }\ c\ \left({1\ +\ r}\right)\cr}
\end{equation}
\end{samepage}

\begin{samepage}
 Now, r can't equal zero, because then the second equation would imply that  ${P}_{\beta }$ is zero for contradiction to our assumptions. Further, r must be real to render a real probability in this equation. Then, straight off we see from first equation that our eigenvalues, probabilities, can't ever be equal to ${P}_{\alpha }$ (unless, contrary to assumptions, c=0) and they further are distinct.
\end{samepage}

\begin{samepage}
Thence, without loss of information, we multiply both sides of first equation by cr to gain a common left hand side :
\begin{equation}
\matrix{{P}_{e}\ c\ r\ =\ {P}_{\alpha }\ \left({1\ +\ {c}^{2}\ r}\right)\ c\ r\cr
{P}_{e}\ c\ r\ =\ {P}_{\beta }\ c\ \left({1\ +\ r}\right)\cr}
\end{equation}
\end{samepage}

\begin{samepage}
which we eliminate for one quadratic equation in r :
\begin{equation}
{P}_{\alpha }\ \left({1\ +\ {c}^{2}\ r}\right)\ c\ r\ =\ {P}_{\beta }\ c\ \left({1\ +\ r}\right)
\end{equation}
\end{samepage}

\begin{samepage}
We rewrite the quadratic in standard form :
\begin{equation}
{c}^{2}\ {r}^{2}\ +\ \left({1\ -\ {{P}_{\beta } \over {P}_{\alpha }}}\right)\ r\ -\ {{P}_{\beta } \over {P}_{\alpha }}\ =\ 0
\end{equation}
\end{samepage}

\begin{samepage}
 and invent an abbreviated notation :
\begin{equation}
{c}^{2}\ {r}^{2}\ +\ \left({1\ -\ Prat}\right)\ r\ -\ Prat\ =\ 0
\end{equation}
\end{samepage}

\begin{samepage}
Our roots are :
\begin{equation}
{r}_{\pm }\ =\ {Prat\ -\ 1\ \pm \ \sqrt {{\left({1\ -\ Prat}\right)}^{2}\ +\ 4\ {c}^{2}\ Prat} \over 2\ {c}^{2}}
\end{equation}
\end{samepage}

\begin{samepage}
Verify, from above roots, expected orthagonality (Hermitian operator of distinct eigenvalues) :
\begin{equation}
\matrix{<{e}_{2}|{e}_{1}>\ \cr
=\ \left({<\alpha \rm |\ +\ c\ {r}_{2}\ <\beta \rm |}\right)\ \left({|\alpha \rm >\ +\ c\ {r}_{1}\ |\beta \rm >}\right)\cr
=\ \left({\matrix{\ \left({<\alpha \rm |\alpha \rm >\ +\ c\ {r}_{1}\ <\alpha \rm |\beta \rm >}\right)\cr
+\ c\ {r}_{2}\ \left({<\beta \rm |\alpha \rm >\ +\ c\ {r}_{1}\ <\beta \rm |\beta \rm >}\right)\cr}}\right)\cr
=\ \left({\ \left({1\ +\ c\ {r}_{1}\ c}\right)\ +\ c\ {r}_{2}\ \left({c\ +\ c\ {r}_{1}\ 1}\right)}\right)\ \cr
=\ 1\ +\ \left({{r}_{1}\ +\ {r}_{2}}\right)\ {c}^{2}\ +\ {r}_{1}\ {r}_{2}\ {c}^{2}\cr
=\ 1\ +\ \left({Prat\ -\ 1}\right)\ -\ Prat\ =\ 0\cr}
\end{equation}
\end{samepage}

\begin{samepage}
Suppose the degenerate case (for eigenvalues= 1/2), then, from the first equation of the system from the eigenvector condition, we have root degeneracy equating the radicand to zero :
\begin{equation}
{\left({1\ -\ Prat}\right)}^{2}\ +\ 4\ {c}^{2}\ Prat\ =\ 0
\end{equation}
\end{samepage}

\begin{samepage}
c was assumed positive WOLOG, so solving c we get one root :
\begin{equation}
c\ =\ {1 \over 2}\ \sqrt {-\ {1 \over Prat}\ -\ Prat\ +\ 2}
\end{equation}
\end{samepage}

\begin{samepage}
Now, only were radicand above positive definite would c be positive definite. Differentiation shows one extremum (Prat = -1 extraneous as not trace normalized).
\begin{equation}
\matrix{y\ =\ -\ {1 \over Prat}\ -\ Prat\ +\ 2\cr
0\ =\ {\partial \rm y \over \partial \rm Prat}\ =\ {1 \over {Prat}^{2}}\ -\ 1\cr
Prat\ =\ 1\cr
then\ plugging\ in\cr
y\ =\ 0\ and\ c\ =\ 0\cr}
\end{equation}
\end{samepage}

\begin{samepage}
We observe radicand = - ° for Prat = 0 or °, so that extremum is a maximum wherein c = 0.
\end{samepage}

\section{Conclusion :}
Spectral decomposition gives probability interpretation of mixture of projection operators. Degeneracy of such ensemble solution is only afforded in special case where projectors are both orthogonal and precisely equal probability.
\section{Suggestions for further reading :}
 I have constructed herein a formal, hopefully reassuring, brief on degeneracy of density matrix versus ensemble superselection.

 Such formal briefs, while surveyable and publishable, can't really get the flavor of the story, the context, the references, the bigger perspective.

 So, I invite the reader to a perspective that not merely from density matrix clarifies the formal probability definition, but exemplifies, in measurement apparatus, probability's wisely designed implementation. 

 For example, I have since my paper come across a delightful paper by Albrecht \cite{a93q} who helped me relate the terms "macroscopic measurement apparatus" to this degeneracy. He finds increasing a measuring instrument's "size" (Hilbert subspace dimensionality) could make superselection insensitive to a "degree" of degeneracy.

 Albrecht points out another feature important to implement correctly. Measurement interaction should insure (QND) stability of the measured factors in the products that superpose in the Schmidt correlated states.

 Consider this stability.

 While Scully in an invited tutorial paper \cite{ScullySticky70} connects different people's notions of the coherent (classical superselected) states, Zurek's pointer basis \cite{QND} speaks of this stability against environmental measurement decoherence of these same states \cite{codecoh}.

 Finally, we "need the story" recorded by this stability! 

 Although reconstruction formidable \cite{Humpty88}, the essential quantum reality of the observed system before an experiment does not have to be disturbed or held as specious by such a well designed apparatus, in fact, throughout (both before and after) the experiment, the dynamics of the quantum reality enables that experiment's design.

\section {Acknowledgements :}

I would like to acknowledge long correspondence with Ulrich Gerlach about his class notes needed to help me attempt to understand HawkingÕs papers, ``Origin of Structure'' and ``Origin of Time Asymmetry'', and thereby WKB time, a term coined by Dr. C. Kiefer. Dr. C. Kiefer long ago corresponded with me about decoherence and time. I am grateful for an even longer correspondence with H. D. Zeh on many other notions of decoherence and for his preprints and suggested readings. I am most happy to have met  Seth Lloyd who encouraged me at Texas InstrumentsÕ Central Research Laboratory. I had the pleasure to hear and question Zurek in a small group at SMU.   Dr. Cy Cantrell reminded me a density matrix was a Hermitian operator and therefore must have orthogonal eigenstates. He pointed me to very interesting literature on coherence and Pi pulses. Dr. Wolfgang Rindler, through much discussion, helped me to get my ideas somewhat clearer.

\end{document}